\documentclass[english,aps,prb,twocolumn,secnumarabic,amsmath,amssymb,superscriptaddress]{revtex4}
\usepackage{color}
\usepackage{soul}
\usepackage{amsmath,amstext}
\usepackage{amssymb}
\usepackage{graphicx}
\usepackage{esint}
\def \beq {\begin{equation}}
\def \eeq {\end{equation}}
\def \ba {\begin{eqnarray}}
\def \ea {\end{eqnarray}}
\newcommand{\upp}{\hspace{-0.2 pt}\uparrow}
\newcommand{\downn}{\hspace{-0.2 pt}\downarrow}

\newcommand{\mean}[1]{\langle#1\rangle}

\newcommand{\opd}[1]{\hat{#1}^{\dagger}}
\newcommand{\op}[1]{\hat{#1}}

\def\ket#1{\left| #1\right>}
\def\bra#1{\left< #1\right|}

\makeatletter
\@ifundefined{textcolor}{}
{%
 \definecolor{BLACK}{gray}{0}
 \definecolor{WHITE}{gray}{1}
 \definecolor{RED}{rgb}{1,0,0}
 \definecolor{GREEN}{rgb}{0,1,0}
 \definecolor{BLUE}{rgb}{0,0,1}
 \definecolor{CYAN}{cmyk}{1,0,0,0}
 \definecolor{MAGENTA}{cmyk}{0,1,0,0}
 \definecolor{YELLOW}{cmyk}{0,0,1,0}
}

\usepackage{babel}

\begin{document}
\title{ \textcolor{black} {Chemical potential for light by parametric coupling}}

\author{M. Hafezi}
\affiliation{Joint Quantum Institute, NIST/University of Maryland, College Park,
MD 20742 }
\affiliation{Department of Electrical and Computer Engineering and IREAP, University of Maryland, College Park MD 20742, USA}
\author{ P. Adhikari}
\affiliation{Joint Quantum Institute, NIST/University of Maryland, College Park,
MD 20742, USA }
\author{J. M. Taylor}
\affiliation{Joint Quantum Institute, NIST/University of Maryland, College Park,
MD 20742, USA }

\begin{abstract}
\textcolor{black}{Usually photons are not conserved in their interaction with matter. Consequently, for the thermodynamics of photons,} while we have a concept of temperature for energy conservation, there is no equivalent chemical potential for particle number conservation.
However, the notion of a chemical potential is crucial in understanding a wide variety of single- and many-body effects, from transport in conductors and semiconductors to phase transitions in electronic and atomic systems. Here we show how a direct modification of the system-bath coupling via parametric oscillation creates an effective chemical potential for photons even in the thermodynamic limit. \textcolor{black}{In particular, we show that the photonic system equilibrates to the temperature of the bath, with a tunable chemical potential that is set by the frequency of the parametric coupler.} Specific implementations, using circuit-QED or optomechanics, are feasible using current technologies, and we show a detailed example  demonstrating the emergence of Mott insulator-superfluid transition in a lattice of nonlinear oscillators. Our approach paves the way for quantum simulation, quantum sources, and even electron-like circuits with light.
\end{abstract}
\maketitle

\section{Introduction}

The study of the thermodynamics of photons dates back to Planck \cite{Planck:1914vs}. Investigating  blackbody radiation, he realized photons decay due to absorption into walls of their container, and therefore, no chemical potential appeared in his expression, in contrast to Gibbs's thermodynamic expressions for other particles using the grand canonical ensemble. Later, it was understood that in the absence of absorbing walls, photon can acquire non-zero chemical potential, e.g. photon emission in semiconductors (LED) \cite{Wurfel:1982wi}, and thus the useful concept of chemical potential can start to be applied to these systems \cite{Ries:1991,Herrman:2005,Job:2006}. Moreover, if photons are confined in a cavity and coupled to excitons, they form polaritons which also can thermalize \cite{Keeling:2007gn,Eastham:2001gb,Carusotto:2012tl}. More recently, it was shown that photons can thermalize with a non-zero chemical potential and form a Bose-Einstein condensate \cite{Klaers:2010bi,Sun:2012,Klaers:2012,snoke2013} when interacting with a nonlinear medium.  However, finding a general solution to creating a chemical potential for light remains an open problem~\cite{yukalov:2012}.

 At the same time, photons provide an intriguing quantum degree of freedom for implementing quantum simulators \cite{Angelakis:2007,Greentree:2006,Hartmann:2006,Chang:2008,Hafezi:2012ku,Carusotto:2009} and observing quantum phases of matter~\cite{Carusotto:2012tl}. In quantum simulation, one develops a quantum system with a controlled, known Hamiltonian, enabling simulation of problems that are exponentially difficult on a classical computer. This new paradigm covers a wide range of problems from chemistry \cite{Kassal:2008vz} and quantum field theories \cite{Jordan:2012kn} to strongly correlated electron systems, such as High-$T_c$ superconductors \cite{Buluta:2009ii}. \textcolor{black} {Recently, several theoretical works have shown that photonic systems can have non-trivial photonic states\cite{Feng:1998,Braak:2011} and even many-body effects with zero chemical potential~\cite{Schiro:2012, Zheng:2011,Schiro:2013,Henriet:2014eh}. In the presence of strong nonlinearity photonic system can exhibit blockade effect \cite{Kimble:2005, Englund:2007,Hoffman:2011vo} which can fix the number of photons in the steady-state. In particular, it was recently shown that under specific conditions (flat-band models and with an incompressibility at a certain particle number), photonic systems can be stabilized by single-photon pumping and parametric drive \cite{Kapit:2014dp}.  However, many phenomena that are interesting from a quantum simulation perspective involve thermalization in systems with a \textit{controllable} chemical potential, as a key parameter in phase diagrams. Both are absent for photons.}

\textcolor{black}{Here, we propose a parametric scheme to address the issue of chemical potential and thermalization in photonic systems, extending preliminary concepts~\cite{Weitz:2013} and developing simpler approaches than current theory~\cite{Stoof:2013,Sobyanin:2013,Kirton:2013}. In particular, by parametrically coupling a photonic system to a thermal bath, we show that a photonic system can equilibrate to the temperature of the bath, with a tunable chemical potential given by the frequency of the parametric coupler. Therefore, this scheme makes it possible to control both the temperature and the chemical potential of a photonic system.} We apply our scheme to two platforms, circuit-QED and optomechanical systems, where recent and spectacular progress has been made in controlling and using them in a few quanta regime.  Finally, we conclude by considering how a photonic lattice implementing a Bose-Hubbard model can be driven through the Mott insulator-superfluid (MI-SF) transition~\cite{Fisher:1989vs} using this approach even in the presence of finite dissipation.

\section{Parametric thermalization}

We can understand thermalization via a system-bath picture, where the system of choice with Hamiltonian $H_S$ is coupled via $\lambda H_{SB}$ to a bath with Hamiltonian $H_B$ and initial state $\rho_B \propto \exp(-\beta H_B)$~\cite{Subas:2012fg,Temme:2013fu}.  \textcolor{black}{Our scheme will follow this approach with one small modification: replace the coupling with a parametric coupling via $\lambda \rightarrow 2 \lambda \cos(\omega_p t)$, where $\omega_p$ is the angular frequency at which the coupling is modulated. Therefore, the system-bath Hamiltonian takes the form ($\hbar=1$),}
\beq 
H = H_S + 2\lambda \cos(\omega_p t)H_{SB} + H_B\ . \label{e:ham} 
\eeq
again with initial conditions $\rho_B \propto \exp(-\beta H_B)$. We assume that parametric drive can be characterized by a classical field which can not be depleted. The parametric coupling will enable up- and down-conversion of bath excitations to photons, which will lead to a controlled chemical potentials.

\begin{figure}
\includegraphics[width=0.45\textwidth]{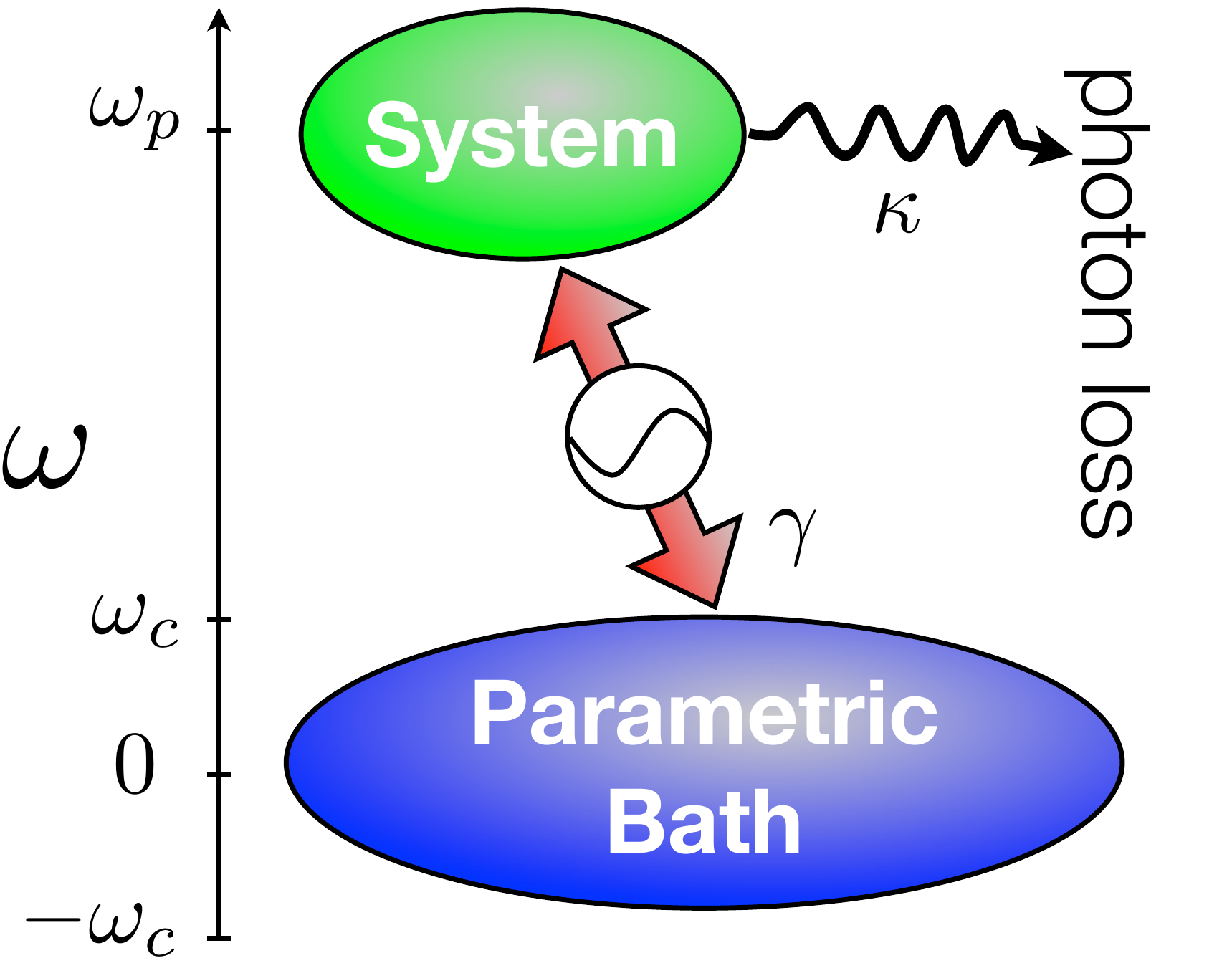}\caption{(a)  Thermal bath with modes $\hat{b}_j$ and response functions with a cutoff $\nu_c$ can be parametrically coupled to a higher frequency (optical) system with modes $\hat{a}_j$ near the frequency $\nu_p$.  Additional loss via the high frequency bath can lead transport from the parametric bath through the system to the high frequency bath.  }
\end{figure}

To see this explicitly, we will assume that $H_{SB}$ is bi-linear, of the form
\beq 
H_{SB} = \sum_j ( \hat{a}_ j + \hat{a}^\dagger_j ) B_ j 
\label{eq:linear_coupling}
\eeq
where $\hat{B}_j$ is a bath operator and there exists $\hat{a}_j$, $n_j$ such that $[\hat{a}_j , n_j ] = \hat{a}_j$, as occurs naturally for photons. This property defines particle numbers $n_j$ and total particle number $\hat N = \sum_j n_j$.  

\textcolor{black}{Let us consider what happens when the energy scales of the bath and system are small compared to $\omega_p$. Specifically, we assume that the system has a low frequency cutoff, and the bath has a low-high frequency cutoff $\nu_c$}. Furthermore, we will decompose $H_S$ into $H'_S +H_{S,\perp}$ where $H_{S,\perp}$ includes all terms that do not commute with the total number of particles in the photonic system, given by $\hat{N}=\sum_j \hat{a}^\dagger_j \hat{a}_j$. \textcolor{black}{Therefore, $H'_S$ is the part of the Hamiltonian that conserves the total number of particles.} In this regime, we move to a rotating frame with the unitary transformation $U = \exp(-i t\omega_p \hat{N})$. The transformed system Hamiltonian becomes
\beq 
U^\dagger H_S U - iU^\dagger \dot{U} \approx H'_S -  \omega_p \hat{N},
\eeq
where we have neglected $U^\dagger H_{S,\perp}U$ by making the rotating wave approximation (RWA), requiring $||H_{S,\perp}|| \ll  \omega_p$.

Meanwhile, the bath Hamiltonian remains the same, while the system bath coupling terms become
\beq 
\left[ \hat{a}_j +\hat{a}^\dagger_j +(e^{-2i\omega_p t} \hat{a}_j +e^{2 i \omega_p t} \hat{a}^\dagger_j)\right]B_j 
\approx \left[\hat{a}_j +\hat{a}^\dagger_j\right]\hat{B}_j 
\eeq
The key approximation is again the RWA to neglect $e^{2 i \omega_p t} \hat{a}^\dagger_j$-type terms, consistent for a bath whose two-point bath correlation function $\mean{B_i(t+\tau) B_j(t)}$ has a cutoff frequency $\nu_c < \omega_p$.  This provides our definition of a low frequency bath for this paper, with $H'_{SB} \equiv \sum_j \left[\hat{a}_j +\hat{a}^\dagger_j\right]\hat{B}_j$ the system-bath coupling in the RWA.

Through this set of transformations, and the rotating wave approximation, we have a new system-bath Hamiltonian which takes the traditional form
\beq 
H = H'_S - \mu \hat{N} + \lambda H'_{SB} + H_B 
\eeq
where we identity $\mu \equiv  \omega_p$ as the chemical potential.
For weak coupling $\lambda$ and an infinite bath at inverse temperature $\beta$, we expect the system to thermalize in the long-time limit to a density matrix
\beq 
\rho \approx \exp\left[-\beta(H'_S -\mu\hat{N})\right], \label{eq:rho_thermal} 
\eeq 
i.e., the distribution is exactly that of the grand canonical ensemble. 

 \textcolor{black}{The key idea of our approach is to parametrically couple a low-temperature, low frequency
bath to a set of high frequency modes. The parametric coupler up-converts bath excitations
to photons and down-converts photons to bath excitations, as shown
in Fig.~1.  This leads to thermalization of photons, as long as
the bath thermalization rate and the coupling rate between the bath
and photons is faster than other photonic decay rates.} 

\section{Implementations}

\begin{figure}[htbp]
\centering 
\includegraphics[width=0.45\textwidth]{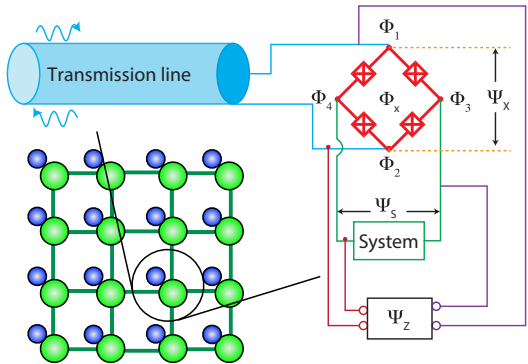} 
\caption{Coupled array of nonlinear microwave cavities provides a potential quantum simulator, where individual elements' parametric coupling to a bath provides chemical potential.  The inset shows the bath coupler implementation suggested in the text using circuit QED.  Specifically, a transmission line is coupled to the mode $\Psi_{X}$ of the coupler. The system is connected to the mode $\Psi_{S}$. The mode $\Psi_{Z} = \lambda \Phi_{0}  \cos \omega_{p} t$ is driven harmonically at frequency $\omega_{p}$ and provides the up- and down-conversion necessary for particle and hole exchange with the bath. }
\label{Wheatstone}
\end{figure}

Now we show that such a scheme, which provides both thermalization and a finite chemical potential for photons, can be implemented in circuit-QED systems for microwave domain photons and using optomechanics for optical domain photons. Following the Caldeira-Leggett model, in the context of circuits  \cite{Devoret:1995,Clerk:2010p679}, we consider the bath to be a collection of transmission lines which can be described by a quasi-continuum of harmonic oscillators. The bath Hamiltonian is given by
\ba
H_B=\sum_{\nu}   \omega_{\nu} \left( \opd{b}_{\nu} \op{b}_{\nu} + \frac{1}{2} \right),
\ea
where $\opd{b}_{\nu}$ is the creation operator of an electromagnetic field quantum at mode $\nu$ with frequency $\omega_{\nu}$.  We assume that the transmission lines are in thermal equilibrium, and thus, $\langle \opd{b}_{\nu} \op{b}_{\nu'}\rangle=\frac{1}{e^{\omega_{\nu}/k_{B}T}-1}\delta_{\nu,\nu'}$. We consider that each mode of the photonic system is coupled to the bath using non-degenerate parametric amplifiers, through three-wave mixing. While many configurations can implement this concept \cite{ZakkaBajjani:2011in}, we focus on the conceptually cleanest case: a Josephson parametric amplifier in a Wheatstone bridge configuration \cite{Bergeal:2010iu}, as depicted in Fig.~\ref{Wheatstone}. 

Examining the details of the JJ-Wheatstone parametric coupler, we assume that each junction has a large area, and hence, a large capacitance, so that its charging energy can be ignored. In this approximation, the energy $U$ of the JJ-Wheatstone bridge is~\cite{bergeal.2010.296--302}
\begin{align}
-4E_{J} &{} \Big[\cos \left(\frac{\Phi_{x}}{4 \varphi_{0}} \right) \cos \left(\frac{\Psi_{X}}{2  \varphi_{0}} \right) \cos \left(\frac{\Psi_{S}}{  2\varphi_{0}} \right) \cos \left(\frac{\Psi_{Z}}{2  \varphi_{0}} \right)  \nonumber \\
&{} + \sin \left(\frac{\Phi_{x}}{4  \varphi_{0}} \right) \sin \left(\frac{\Psi_{X}}{2  \varphi_{0}} \right) \sin \left(\frac{\Psi_{S}}{2  \varphi_{0}} \right)\sin \left(\frac{\Psi_{Z}}{2  \varphi_{0}} \right) \Big] \nonumber
\end{align}
where we have taken all four JJ's to have the same $E_J$, and $\varphi_{0} = \Phi_{0}/(2\pi)$, $\Phi_{0} = h/(2e)$ being the superconducting flux quantum. Setting $\Phi_{x} = \Phi_{0}/2$ by choice of flux bias, and assuming the mode intensities $\Psi_{X}, \Psi_{S}, \Psi_{Z} \ll \Phi_{0}$, consistent with moderate to low characteristic impedance circuits, we can expand $U$ in $\psi_{i} = \Psi_{i} / \Phi_{0}$, $i \in \{X,S,Z\}$ to third order~\cite{abdo2013TWM}:
\ba
U = -2 \sqrt{2} E_{J} + M \left(\psi_{X}^2 + \psi_{S}^2 + \psi_{Z}^2 \right) + g \psi_{X} \psi_{S} \psi_{Z}\ .
\ea
Here $M = \sqrt{2} E_{J} \pi^2$ and $g = -2\sqrt{2} E_{J} \pi^3$. 

 \textcolor{black} {A transmission line is connected inductively to the coupler mode $\Psi_{X} = \Phi_{1} - \Phi_{2}$ through inductance $L_{1}$. The modes, assumed to be in thermal equilibrium at a temperature $T$, act as a bath. The (microwave) photonic system is coupled to the mode $\Psi_{S} = \Phi_{4} - \Phi_{3}$ while the mode $\Psi_{Z} = \Phi_{1} - \Phi_{3} + \Phi_{2} - \Phi_{4}$ is externally modulated as $\Psi_{Z} = \lambda \Phi_{0}  \cos (\omega_{p} t+\phi)$, where $\lambda$ is the dimensionless amplitude of the modulation and controls the system-parametric bath coupling strength.}

 \textcolor{black} {Let $C_{l} = C L$ be the capacitance of the transmission line, with $C$ being its capacitance per unit length. Because of the presence of the transmission line, $\Psi_{X} = \sum_{\nu} \psi_{\nu} \sqrt{\frac{1}{2 C_{l} \nu}} (b_{\nu} + b_{\nu}^{\dagger})$. Here, $\psi_{\nu}$ is a dimensionless parameter that depends on the boundary conditions at $z=L$.  For our particular coupling -- current-flux -- we expect $\psi_{\nu} \sim \sin(k_\nu L)$ and, in the weak coupling limit, $\psi_{\nu } \propto \nu$. Ignoring coupling between different transmission line  modes, the system Hamiltonian is
\ba
H_S + H_B +  \lambda \cos (\omega_{p} t+\phi)  \sum_{\nu}  h_{\nu}  (b_{\nu} + b_{\nu}^{\dagger}) \Psi_{S},
\ea
where $h_{\nu} = \frac{g}{\Phi_{0}^{2}} \psi_{\nu} \sqrt{\frac{1}{2 C_{l} \nu}}$.
This then directly produces our model Hamiltonian for generating a chemical potential, where the density of states $J(\nu) = h(\nu)^2 \rho(\nu) \propto \nu$, i.e., an Ohmic bath~\cite{Clerk:2010p679}.}

For the optical domain, we need a different parametric process.  A convenient one is the optomechanical coupling between motion of a mirror and the frequency of light in a cavity formed by the mirror.  This example case has been worked in partial detail in Ref.~\cite{Weitz:2013}.  The key idea is for a pump field to take the radiation pressure coupling $a^\dag a x$ to a fast oscillating coupling via $a \rightarrow a + \alpha e^{-i \omega_p t}$, producing a parametric coupling to the phonon ``bath'' with frequency $\omega_p$.  The details and benefits of the optomechanical approach will be considered in a separate work. 

Note that in any experimental implementation, one needs to filter out the pump photons from the signal system photons. This can be easily achieved by using different polarization or spatial modes of the photonic system on each site. Alternatively, in certain schemes, one can reject the pump by frequency filtering. For example, in the Mott insulator case, discussed later in this article, the pump has higher frequency than the prepared Mott state, and therefore, the pump can be filtered out by frequency selection.

\section{Bath Discussion}
We now examine our assumption of a cutoff in the bath degrees of freedom, as well as a strictly parametric system-bath coupling. For simplicity, we divide the bath modes into three, independent sets of modes, and consider coupling to a single system mode $a$. Given a parametric coupling at frequency $\nu$, the low frequency modes of the bath, $b_j$, are defined as those with natural resonance frequencies $\omega_j \leq \nu/2$.  The `natural' modes, $c_j$, are those with frequencies $\nu/2 < \omega_j \leq 3 \nu/2$. The `doubly rotating' modes, $d_j$, are those with frequencies $\omega_j > 3 \nu/2$.  Thus, the more general system-bath interaction is
\begin{eqnarray}
H_{SB} &=& \left[A + \lambda \cos(\nu t)\right] (a + a^\dag) \\
&\times& \sum_j f_j (b_j + b_j^\dag) + g_j (c_j + c_j^\dag) + h_j (d_j + d_j^\dag) \nonumber
\end{eqnarray}

We now move to an appropriate rotating frame, with $a \rightarrow a e^{-i \nu t}$, $b_j \rightarrow b_j$, $c_j \rightarrow c_j e^{-i \nu t}$, and $d_j \rightarrow d_j e^{-2 i \nu t}$.  With the assumption of weak coupling ($A, \lambda$ small), we look at the rotating wave approximation for the different couplings:
\begin{align}
H_{SB,b} &= \left[A + \lambda \cos(\nu t) \right] \left(a e^{-i \nu t} + a^\dag e^{i \nu t}\right)\sum_j f_j (b_j + b_j^\dag) \nonumber \\
& \rightarrow  \frac{1}{2} \lambda (a + a^\dag) \sum_j f_j (b_j + b_j^\dag) \\
H_{SB,c} &= \left[A + \lambda \cos(\nu t) \right] \nonumber \\
&\times \left(a e^{-i \nu t} + a^\dag e^{i \nu t}\right)\sum_j g_j (c_j e^{-i \nu t} + c_j^\dag e^{i \nu t}) \\
& \rightarrow  A \sum_j g_j (a^\dag c_j + c_j^\dag a) \\
H_{SB,d} &= \left[A + \lambda \cos(\nu t) \right] \left(a e^{-i \nu t} + a^\dag e^{i \nu t}\right) \nonumber \\
& \times \sum_j h_j (d_j e^{-2 i \nu t} + d_j^\dag e^{2 i \nu t}) \\
& \rightarrow  \frac{1}{2}  \lambda \sum_j h_j (a^\dag d_j + d_j^\dag a) 
\end{align}
By breaking up the bath into three different frequency regions, we see that the `natural' and `doubly-rotating' frequency regions both lead to a system-bath interaction of the quantum optics type, i.e., that of a beam splitter interaction $a^\dag c + c^\dag a$. For these portions of the system-bath interaction, we may then proceed in deriving the master equation in the usual way~\cite{Gardiner:2011p47031,HeinzPeterBreuer:2011wp}, and find, in appropriate limits, a decay of excitations of $a$ at a rate $\kappa \sim A^2 |g_j|^2 \rho(\nu) + \frac{1}{4} \lambda^2 |h_j|^2 \rho(2 \nu)$, where $\rho(\nu)$ is the bath density of states near the parametric modulation frequency $\nu$.

We now investigate the remaining portion of the system-bath interaction in a specific setting, to illustrate the emergence of a Mott insulator-superfluid transition in a photonic lattice.

\section{Lattice model and master equation}

We consider now what happens to a lattice of coupled, interacting photonic resonators, coupled to both a parametric bath at inverse temperature $\beta$ and nominal coupling rate $\gamma$ and to a high frequency (loss) bath with loss rate $\kappa$.  For simplicity, we consider only strong on-site repulsion $U$, and have for the conservative parts of the evolution, a Bose-Hubbard Hamiltonian~\cite{Fisher:1989vs} in the rotating frame:
\begin{align*}
H_S =&{} H_0 + H_J, \ \textrm{with}\\ 
H_0 =&{} \sum_i \left[ \frac{U}{2} n_i (n_i - 1) - \mu n_i \right] \ \textrm{and} \\
H_J =&{} -J \sum_{\mean{ij}} a_i^\dag a_j,
\end{align*}
\textcolor{black}{where $J$ is the tunneling rate between adjacent sites.}

We explicitly derive the master equation for the system, using the usual prescription: first, move to the interaction picture with respect to $H_S + H_B$, where $H_B$ is the bath Hamiltonian and the $\lambda$ prefactor in the system-bath coupling will be a perturbative parameter.  We can write the evolution equation for short times $\tau$ as
\[
\dot \rho_I(\tau) = -i \lambda [H_{SB}(\tau), \rho_I(0) ] - \lambda^2 \int_0^\tau [H_{SB}(\tau),[H_{SB}(t),\rho_I(t)]] dt
\]
with $H_{SB}(t) = \sum_j B_j(t) x_j(t)$ the system-bath coupling in the interaction picture, writing $x_j(t) = a_j(t) + a_j^\dag(t)$.

Now we make the Born and Markov approximations.  That is, we replace $\rho_I(t)$ with $\rho_S(\tau) \otimes \rho_B$.  Here $\rho_B$ is the bath density matrix which will be time-translation invariant for an infinite bath, and is independent of $\rho_S$ with $\mean{B_i} \equiv \textrm{Tr}_B[B_i \rho_B] = 0$ for all bath operators coupled to the system.  From these two approximations, we can trace over the bath and recover the master equation (in the interaction picture)
\begin{align}
\dot \rho_S(\tau) =&{} -\sum_{ij} \int_{0}^\infty S_{ij}(t) [x_i(\tau) x_j(\tau-t) \rho_s - x_i(\tau) \rho_S x_j(\tau-t)] \nonumber \\
&{} \ + S_{ij}(-t) [\rho_s x_j(\tau-t) x_i(\tau)   - x_j(\tau-t) \rho_S x_i(\tau)] dt
\end{align}
with $S_{ij}(t) = \lambda^2 \textrm{Tr}_B[B_i(t) B_j(0)]$ the bath correlation function and where, by taking the initial integration point to $-\infty$, we have assumed that bath correlations decay faster than the effective damping they induce -- consistent with the Markov approximation.

At this point, we wish to develop a time-local master equation.  We express $x_j(t)$ in the energy eigenbasis of $H_S$, with states $\ket{k}$ and energies $\epsilon_k$ and an ordering in energy such that $k' > k \rightarrow \omega_{k'k} \equiv \epsilon_{k'} - \epsilon_{k} \geq 0$.  Then \begin{equation}
c_j(t) = \sum_{l > k} e^{-i \omega_{lk} t} x_{j,kl} \ket{k} \bra{l}\ ,
\end{equation}
formally defines an operator that reduces or keeps constant the energy, and $x_j(t) = c_j(t) + c_j^\dag(t) + x_0$, with the last term time-independent and neglected in what follows.

\textcolor{black}{Taking independent, Ohmic baths for each coupling term, we have
\begin{equation}
S_{ij}(t)
=  \frac{\delta_{ij}}{\pi} \int_0^\infty d\nu J(\nu) \left[ (N_{th}(\nu)+1) e^{-i \nu t} + N_{th}(\nu) e^{i \nu t} \right]
\end{equation}
with the effective spectral density  $J(\nu) = \nu e^{-\nu/\nu_c}$, $N_{th}(\nu) = 1/[\exp(\beta \nu)-1]$, where $\beta$ is the inverse temperature of the parametric bath and $\nu_c \gg U$ is a high frequency cutoff that is irrelevant to the rest of our calculation.  At this point, we get terms in the master equation of the form $S_{ij}(t) (c_i(\tau) c_j(\tau-t) \rho_S)$ and terms of the form $S_{ij}(t) (c_i(\tau) c_j^\dag(\tau-t) \rho_S)$.  The former will have phase evolution at a finite frequency as a function of $\tau$, and will be neglected in a rotating wave approximation.  The latter will also have such terms, except for those with $\omega_{kl} = \omega_{k'l'}$, i.e., energy-degenerate transitions.  Keeping only these transitions immediately takes us to the usual golden rule result: transitions with a positive energy difference $\nu$ occur with a rate $J(\nu) N_{th}(\nu)$ and transitions with a negative energy difference have the rate $J(\nu) [N_{th}(\nu) + 1]$.}

Thus, when the energy levels of the system are well resolved, we can derive a super operator describing both photon loss and coupling to the parametric bath.  Using the commutation of $H_S$ with $N$ (the total photon number), we get transitions from $k$ to $l$ with rates that depend on whether the total photon number of the two states differs by $+1$ or $-1$ as:
\begin{align}
\Gamma^{+}_{k \rightarrow l} &= \gamma \left(N_{th}(|\epsilon_k-\epsilon_l|) + \Theta(\epsilon_k - \epsilon_l) \right)  \sum_i \left| \bra{l} a_i^\dag \ket{k} \right|^2 \label{e:Gammaplus} \\
\Gamma^{-}_{k \rightarrow l} &= \left[\gamma \left(N_{th}(|\epsilon_k-\epsilon_l|) + \Theta(\epsilon_k - \epsilon_l) \right)  + \kappa \right] \sum_i \left| \bra{l} a_i \ket{k} \right|^2  \label{e:Gammaminus} 
\end{align}
where $\gamma = \gamma_0 \frac{|\epsilon_k - \epsilon_l|}{U}$ for the Ohmic bath case, $\gamma_0$ represents the overall strength of the coupling, and $\Theta$ is the Heaviside step function.  We have gone back to the physical couplings $a_i$ rather than the many-body energy lowering operator $c_j$ in order to make clear the special role loss via the high frequency bath plays in Eq.~\ref{e:Gammaminus}.

\begin{figure}
\includegraphics[width=0.45\textwidth]{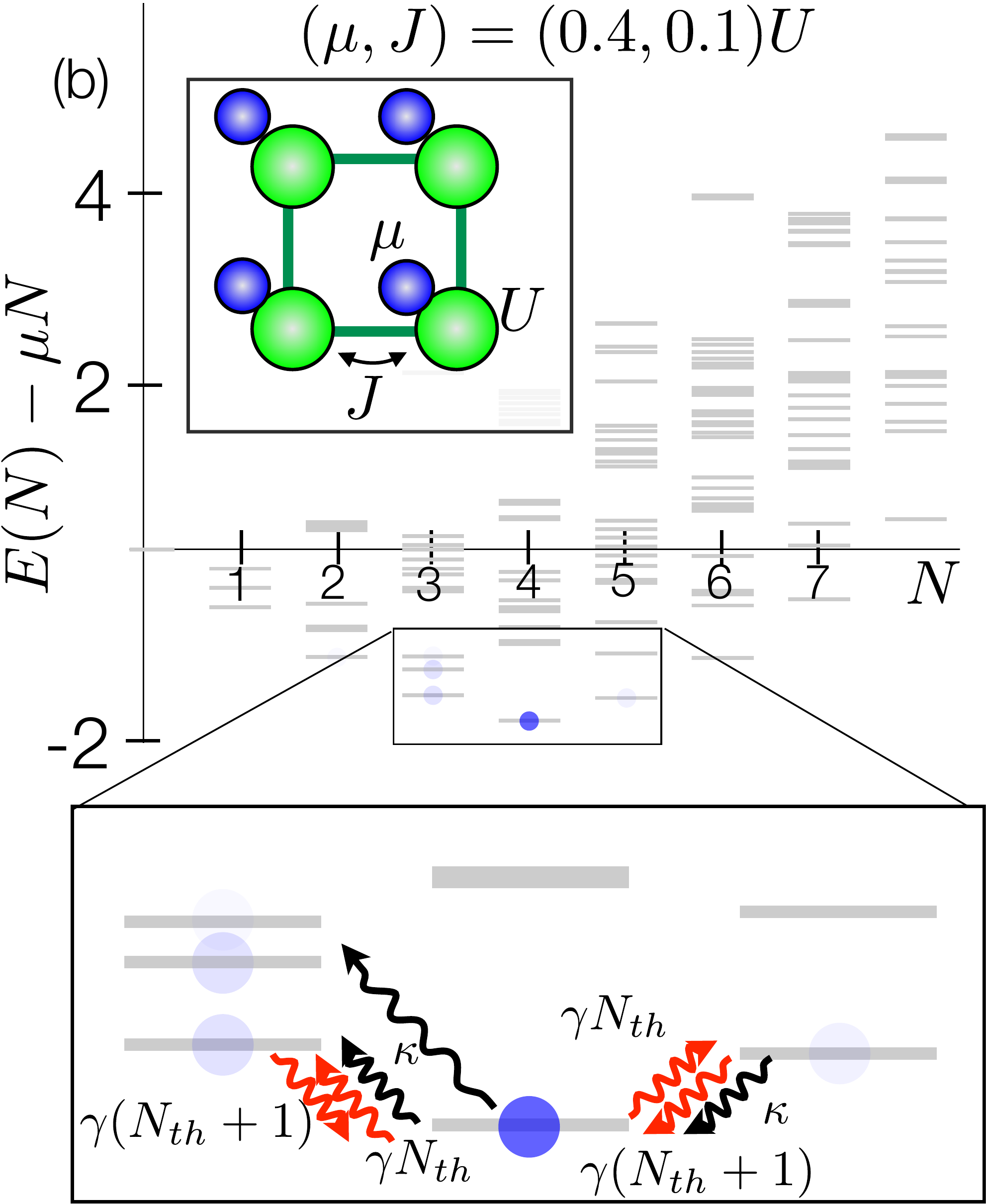}
\caption{\label{f:exact}
Energy eigenstates plotted as a function of energy and total photon number $N$ for an numerical solution of a four site Bose-Hubbard model (shown in the upper inset) with $(J,\mu,\gamma,\kappa) = (0.1,0.4,0.01,0.003) U$ and $\beta = 1/10 U$.  The opacity of the blue dots represent the probability, in steady state, of being in the associated energy eigenstate.  The lower inset shows the region near the ground state in the rotating frame; hole-like excitations (lower $N$) are preferentially filled due to optical loss processes $\kappa$ only reducing particle number.  The relatively high temperature leads to some thermal filling of the first particle excited state.}
\end{figure}

The superoperator takes Lindblad form with these rates leading to a rate equation in the energy eigenbasis.  Solving this numerically for a case of four coupled sites (Fig.~\ref{f:exact}), we can immediately see an intuitive understanding of the two types of decay processes.   The first type, which increases photon number, corresponds to the decay of holes (if the energy of the higher photon number state is lower in the rotating frame) or the creation of particles (if otherwise).  The second type decreases photon number, and includes both creation of holes via loss and via the parametric bath; consequently, we expect a greater rate for the second process, which will lead to a particle-hole temperature asymmetry as shown below.  The simulations themselves correspond to fixing a maximum total particle number per site, finding the eigenenergies of the dissipation-free model, calculating the decay rates in Eqs.~(\ref{e:Gammaplus}) and (\ref{e:Gammaminus}), determining the steady state of the master equation, and for that steady state, finding the probability of each state (shown in the inset to Fig.~\ref{f:exact}), and estimating the Mandel \textcolor{black} {$Q=\frac{\langle n^2\rangle-\langle n\rangle^2-\langle n\rangle}{\langle n\rangle}$}  parameter and the average hopping $\mean{a} \equiv \sqrt{|\mean{a_i^\dag a_j}|}$ (shown in Fig.~\ref{f:meanfield}).

\section{Strong interaction expansion}

We now take a simpler form of the superoperator describing both photon loss and coupling in the case of a single resonator site ($J=0$) to get an analytical handle on the process.  That is, we evaluate Eqs.~(\ref{e:Gammaplus}) and (\ref{e:Gammaminus}) in the single site case.
%
%
Specifically, defining $E_0(n) =  \frac{U}{2} n (n - 1) - \mu n$, the sign of $\Delta E(n) = E_0(n+1) - E_0(n) = n U - \mu$ determines both the direction of decay and the thermal bosonic enhancement factor $N_{th}(|\Delta E(n)|)$.  Thus $\Gamma^{+}_{n \rightarrow n+1} = \gamma f_+(n), \Gamma^{-}_{n+1 \rightarrow n} = (n+1) \kappa + \gamma f_-(n)$ with
\begin{align}
f_+(n) &= (n+1) \left[ N_{th}(|\Delta E(n)|) + \Theta(-\Delta E(n)) \right]\ , \\
f_-(n) &= (n+1) \left[ N_{th}(|\Delta E(n)|) + \Theta(\Delta E(n)) \right]\ .
\end{align}
and $\gamma = \gamma_0 |\Delta E(n)|/U$ for the Ohmic bath case.
The change from $N_{th}$ to $N_{th}+1$ that occurs in these two factors with the change in sign of $\Delta E(n)$ arises from having both co- and counter-rotating terms in the system bath coupling.

One consequence of the strong interaction (sometimes called strong coupling in the Mott insulator literature) limit ($J \rightarrow 0$) is an analytical form for the steady state.  Specifically, we recover a form of detailed balance, where the probability of a transition on a site from photon number $n$ to $n+1$ is given by $\gamma f_+(n)$ while the transition from $n+1$ to $n$ is $\gamma f_-(n) + (n+1) \kappa$.  This gives, in steady state, a set of ratios
\begin{align}
\frac{p_1}{p_0} =&{} \frac{ f_+(0)}{f_-(0) + \kappa/\gamma} \\
\frac{p_2}{p_1} =&{} \frac{f_+(1)}{f_-(1) + \kappa/\gamma} \\
\ldots &{}
\end{align}
where the correction from a thermal distribution arises from the term $\kappa/\gamma$, which depends on the energy difference via $\gamma$. We can characterize this for two regimes.  First, when $\Delta E(n)$ is positive (it costs energy to add a photon), we expect the ratio $p_{n+1} / p_{n} = N_{eff}^{(p)} / (N_{eff}^{(p)} + 1)$.  This defines the bosonic occupation as seen by particle addition as 
\[
N_{eff}^{(p)} = \frac{N_{th}(|\Delta E(n)|) }{1 + \kappa/\gamma}
\]
Thus, when particles cost energy, photon loss reduces the effective temperature of the system.

Similarly, when $\Delta E(n)$ is negative, we expect $p_{n+1} / p_{n} = (N_{eff}^{(h)}+1) / N_{eff}^{(h)}$, which defines the bosonic occupation as seen by hole addition:
\[
N_{eff}^{(h)} = \frac{\left[ N_{th}(|\Delta E(n)|) + \kappa/\gamma \right]}{1 - \kappa/\gamma}
\]
Here, photon loss increases the energy, and thus increases the effective temperature of the system.  Furthermore, any hope of a thermal description will necessarily breakdown for $\kappa/\gamma \geq 1$.

%

Having established that a Mott insulator-like phase emerges from the single site picture, we can ask how the asymmetry of particles and holes changes the standard picture of the edges of the Mott lobes, by using a picture of free particles and holes above the $n_0$ particle-per-site Mott state $\ket{\Psi_M} \propto \prod_i (a_i^{\dag})^{n_0} \ket{\textrm{vac}}$, i.e., using small $J$ perturbation theory in the strong interaction limit.  A crucial difference from the standard treatment~\cite{Freericks:1994wt} is the use of an implicit finite lifetime to such excitations due to the coupling to both parametric and high frequency baths.

When $J$ exceeds damping and dephasing, we can no longer use a master equation appropriate to a single site.  Specifically, as we want the parametric bath to resolve the kinetic terms in the Hamiltonian, we require $\gamma [= \gamma_0 (J/U)] \ll J$.  We can, however, characterize the particle or hole occupation for a wave vector $k$ in the dilute limit (where particle hole collisions are neglected) by using our $N_{eff}^{(p[h])}$, and we can ask over what domain of parameter space is the combined occupation of particles and holes small compared to one per site.  Here we rely upon the standard picture of particle and hole energies to order $J^2/U$, neglecting loss-induced changes to the energy differences, consistent with $\kappa \leq \gamma \ll J$.  The energy of a particle(hole) of wave vector $k = 0$ above the Mott state is given by Ref.~\cite{Freericks:1994wt} and reproduced here to order $J^2/U$:
\begin{widetext}
\begin{align}
\Delta E^{(p)} =&{} -z J (n_0 + 1) + n_0 U - \mu + \frac{z J^2}{2 U} n_0 (5 n_0 + 4) - \frac{z^2 J^2}{U} n_0 (n_0 +1) \\
\Delta E^{(h)} =&{} -z J (n_0) - (n_0-1) U + \mu + \frac{z J^2}{2 U} (n_0 +1) (5 n_0 + 1) - \frac{z^2 J^2}{U} n_0 (n_0 +1) 
\end{align}
\end{widetext}
where $z$ is the number of nearest neighbors.  

\begin{figure}
\includegraphics[width=0.5\textwidth]{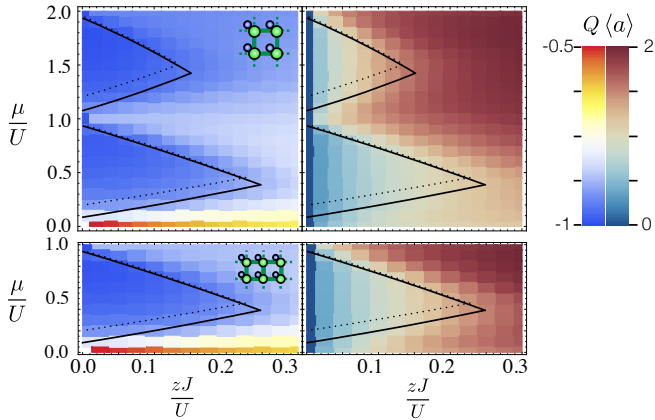}
\caption{\label{f:meanfield}
Numerical results for Mandel $Q$ (left) and coherence $\mean{a} \equiv \sqrt{|\mean{a_i^\dag a_j}|}$ (right) using the four-site (top) and a six-site (bottom) Bose-Hubbard model with periodic boundary conditions.  We assume an Ohmic parametric bath and a flat high frequency (loss) bath.  The Mandel $Q \approx -1$ regions (dark blue, left plot) are the Mott insulator states; at the same time, the finite coherence between sites on the right indicates the emergence of superfluid order (right plot) outside the Mott lobes.  Finite size effects prevent observation of sharp transitions. Here $\gamma_0 = 0.07 U$ and $\kappa = \gamma_0/30$.  Overlaid are the critical values for finite occupation of particles and holes (solid black lines), with the dotted line for a higher value of $\kappa =  \gamma_0/3$.  The asymmetry of particles and holes arises due to preferential hole creation from optical loss.}
\end{figure}

We can then calculate the average particle and hole expectation values including both the parametric bath and the high frequency (photon loss) bath, and find that these lowest energy modes have just $N_{eff}^{(p)}$ and $N_{eff}^{(h)}$ with the above $\Delta E^{(p[h])}$.  The boundary of the phase would then correspond to this effective occupation approaching unity (at which point we may expect a macroscopic occupation of particles and/or holes in the system, taking us far from the Mott state).  This boundary is shown for two different values of $\kappa/\gamma_0$ in Fig.~\ref{f:meanfield}; as $\kappa$ increases, the lobes become asymmetric, consistent with additional hole creation via particle losses.

We now consider what near equilibrium picture can emerge, and in particular focus on a picture with two reservoirs (particles and holes) at different temperatures due to loss into the high frequency bath.  In the limit of $\kappa \rightarrow 0$, we recover the usual picture of an equilibrium system, and get a critical temperature defined as
\[
T_c^{(0)} = \frac{1}{k_B \log 2} \textrm{Min}[\Delta E^{(h)},\Delta E^{(p)}]
\]
However, including the non equilibrium effects, we instead have for the parametric bath temperature the requirement
\begin{equation}
T \leq T_c^{(ne)} =\frac{1}{k_B} \textrm{Min}\left[\frac{\Delta E^{(h)}}{\log \left(\frac{2(\gamma^h-\kappa)}{\gamma^h - 2\kappa} \right)},\frac{\Delta E^{(p)}}{\log \left(\frac{2 \gamma^p+\kappa}{\gamma^p + \kappa} \right)}\right]
\end{equation}
where $\gamma^{h[p]}$ depends on $\Delta E^{(h[p])}$ via $J(\nu)$. 

Further analysis of the particle-hole picture at finite temperature will no doubt elucidate additional physics for this non equilibrium system, following perhaps the efforts of 
Refs.~\cite{CapogrossoSansone:2010p45059,Gupta:2013db}.  In addition, an appropriate mean field theory including modifications of the system-bath coupling could provide insight into the applicability of such theories for describing non-equilibrium systems.

\section{Conclusion}
Providing a robust chemical potential for light allows for classical and quantum systems to access a wide variety of heretofore forbidden domains.  Crucially, our approach allows one to build from well established theoretical tools for non equilibrium problems with chemical potential imbalances, such as occurs in circuits and cold atom systems, rather than the thornier problems associated with driven steady-state systems more typical to the quantum optical domain.  From a quantum simulation perspective, this simplification makes the state preparation problem much more straightforward than existing approaches, and yields a mechanism for robust quantum simulation of condensed matter and chemistry problems with light.  In addition, our parametric coupling scheme has a wide range of potential implementations, all of which are accessible with current technology, and enables a variety of practical applications in the context of non-classical sources in the microwave and optical domain that operate more in analogy to a diode than to a pumped dissipative steady-state system.

We thank S. Girvin, A. Houck, B.~L. Hu, J. Keeling, J. Freericks, M. Devoret, E. Kapit, and P. Zoller for helpful discussions.  Support was provided by the NSF-funded Physics Frontier Center at the JQI and by ARO MURI  Grant No. W911NF0910406.

\section*{Appendix}

As a simple test of these concepts, we implement numerically a model for the intermediate time behavior of a two-level system (qubit) coupled to a bosonic bath via a parametric coupling. The usual picture of quantum Brownian motion~\cite{hanggi:2005} has a set of bath modes coupled linearly through their position variables $x_\omega$ with constant $\tilde g_\omega$ and mass $m_\omega$. This leads to the effective spectral density $J(\omega) = \frac{\rho(\omega) \tilde g_\omega^2}{m_\omega \omega}$ where $\rho(\omega)$ is the density of states. Our goal will be to well approximate such a bath with a discrete set of modes. Before engaging in that, we mention some rescaling of the problem appropriate to simulation. First, we rewrite the system-bath coupling $
\tilde g_\omega x_\omega (a+a^\dag)$ in terms of bath creation and annihilation operators with $x_\omega =\sqrt{\frac{1}{2 m \omega}} (b_\omega + b^\dag_\omega)$. This defines $g_\omega = \tilde g_\omega \sqrt{\frac{1}{2 m \omega}}$ and $J(\omega) = 2\rho(\omega) g_\omega^2$.


Our goal is to approximate the bath such that the correlation function and the commutation relation of the bath are as close to the desired approximate bath as possible. Specifically, for our quantum Brownian motion bath with a cutoff function $f(\omega)$ considered in this work, we assume the time-ordered spectral function used in Sec. 5:
\begin{align}
S(\tau) 
&= \frac{1}{\pi} \int_0^\infty d\nu J(\nu) \left[ (N_{th}(\nu)+1) e^{-i \nu t} + N_{th}(\nu) e^{i \nu t} \right] \nonumber \\
&\approx \sum_{j=1}^N \frac{{\bf w}_j  J(\omega_j)}{\pi f(\omega_j)}  \left[  (N_{th}(\omega_j)+1) e^{-i \omega_j t} + N_{th}(\omega_j) e^{i \omega_j t}
\right]
\end{align}
where the approximation of the integral as a finite sum arises from Gaussian quadrature over $N$ orthogonal polynomials under the function $f(\omega)$ to find the set $\{ \omega_j \}$ and the associated weights ${\bf w}_j$, and $N_{th}(\nu) = 1/[\exp(\beta \nu) - 1]$.  We remark that in the case of the Ohmic bath with exponential cutoff function, the appropriate choice is the Laguerre polynomials.

A simple reinterpretation of this formula is that of a discrete harmonic oscillator (quantum Brownian motion) bath with frequencies $\omega_j$ and a coupling constants 
\[
g_j = \sqrt{\frac{{\bf w}_j J(\omega_j)}{ f(\omega_j) }} .
\]
This approximation is immediately amenable to numerical techniques via direct integration of the Schr\"odinger equation. 

In practice, exponential cutoffs at the relevant frequencies are unlikely as the superconductors work well into the GHz domain for our implementation. Thus, we consider a polynomial cutoff function induced by filtering the Ohmic bath with a low-pass filter, such as a capacitor in parallel with the resistor forming the Ohmic bath for our circuit case. The impedance of this system becomes: $Z=R/(1+i\tau_{RC}\omega)$, where $\tau_{RC}=RC$ is the characteristic time of the RC circuit. Therefore, the real part of the impedance, which appears in the spectral noise \cite{Clerk:2010p679}, leads to a natural modification of the effective spectral density: $J(\omega) \rightarrow J(\omega) \frac{1}{1+ \omega^2 \tau_{RC}^2 }$.  We neglect imaginary contributions to the circuit
by assuming they are renormalized in the system Hamiltonian. We note that for an optomechanical implementation such cutoff functions arise from the cavity Lorentzian and can have a similar functional form -- quadratic suppression at high frequency.

We simulate the following simple case numerically to illustrate our system. Working with the  discrete bath approximation and a photon-blockade-regime cavity with an effective two-level system description with Pauli matrices $\sigma_z$, etc., we write
\beq
H = \omega_0 \sigma_z / 2 + \left[A + \lambda \cos(\mu t) \right] \sum_j g_j \sigma_x (b_j + b_j^\dag) + \omega_j b_j^\dag b_j
\eeq
with $g_j = \sqrt{\frac{\bf{w}_j \omega_j}{1 + \omega_j^2 \tau_{RC}^2}}$.  The parameters $A$ and $\lambda$ represent the relative strength of the regular exponential decay bath and the additional parametric bath terms oscillating at frequency $\mu$.

We take $\tau_{RC} = 4/\omega_0$ and $\omega_{\textrm{cutoff}} = 2.5 \omega_0$, and find that decreasing or increasing $\omega_{\textrm{cutoff}}$ by even a factor of two does not appreciable change the results presented below. We also work in units of time given by $1/\omega_0$. For improved computation speed, we truncate the bath Hilbert space to a maximum of two bosonic excitations, and confirm post-facto that simulations produce only slightly more than one bath excitation, consistent with the truncation.

We first test the purely Ohmic case, taking $\tau_{RC} \rightarrow 0, A = 0.5 / \sqrt{17},$ and $\lambda = 0$ (no parametric bath).  We find exponential decay with a time scale $\gamma_a^{-1} = 16.8(4)$. Furthermore, this decay is well approximated (with around $< 1\%$ errors) up to times $t \gtrsim\{30,50,65, 100\}$ for $N=\{35,50,70,100\}$. The fitted decay rate is independent of $N$ in this range of values, consistent with our approximation scheme.  

\begin{figure}
\includegraphics[width=3.0in]{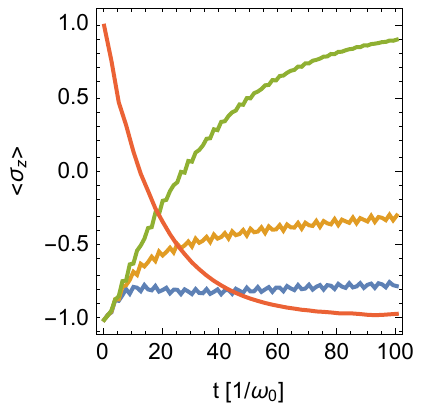}
\caption{
Time-dependent simulation of a two level system with natural frequency $\omega_0$ interacting normally (red) and parametrically (green, yellow, blue) with a high-frequency filtered bath.  For an initial excited state $\mean{\sigma_z}=1$, coupling to the bath leads to exponential decay when the oscillating frequency of the bath is set to zero, as shown in red. However, for an initial ground state $\mean{\sigma_z} = -1$, turning on the parametric coupling to the bath such that it oscillates at frequencies $\mu = \{$ 0.9 (blue), 1.0 (yellow), 1.1 (green) $\} \omega_0$ leads to inversion of the spin when $\mu > \omega_0$, as predicted by our more general theoretical model.
\label{f:decay}}
\end{figure}

We then consider $A=0.5$ with the filter on ($\tau_{RC} \rightarrow 4$), leading to a slightly reduced decay rate due to the non-Ohmic nature of the bath near $\omega_0$ from the cutoff filter. Still, exponential decay is observed over two decades with $\gamma_a^{-1} = 18.7(5)$, and residuals are at the 3\% level or less, largely due to corrections to exponential decay at long times from non-Ohmic bath behavior. The decay from an initially excited state $\ket{\upp}$ into the zero temperature bath is shown in red in Fig.~\ref{f:decay}.

After these simple tests of our model system, we move to the more complicated regime of a parametrically coupled bath. Taking $A=0$ and $\lambda=0.5$, we start the system in the lower-energy spin state, $\ket{\downn}$. We calculate over 100 time units three different values of our chemical potential parameter, $\mu = \{ 0.9, 1.0, 1.1\}$, and plot the resulting $\mean{\sigma_z}$ as a function of time. We expect the system to approach spin up for $\mu > 1$ according to the chemical potential derivation given in the first part of this part, and find these expectations confirmed in this simple numerical experiment (Fig.~\ref{f:decay}).

\end{document}